# Exploring user needs in relation to algorithmically constructed classifications of publications

## A case study

Peter Sjögårde

0000-0003-4442-1360

Health Informatics Centre, Department of Learning, Informatics, Management and Ethics
Karolinska Institutet
Stockholm Sweden peter.sjogarde@ki.se

## ABSTRACT

Algorithmic classification of research publications has been created to study different aspects of research. Such classifications can be used to support information needs in universities for decision making. However, the classifications have foremost been evaluated quantitatively regarding their content, but not qualitatively regarding their feasibility in a specific context. The aim of this study was to explore and evaluate the usefulness of such classifications to users in the context of exploring an emerging research area.

I conducted four interviews with managers of a project aimed to support research and application of artificial intelligence at the Swedish medical university Karolinska Institutet. The interviews focused on the information need of the managers.

To support the project, a classification was created by clustering of publications in a citation network. A cluster map based on this classification was provided to the project leader and one interview focused on the use of the classification in the project in relation to the stated information needs.

The interviews showed that the aim of the project was to improve competence, enhance communication between researchers and develop support structures. Getting an overview of artificial intelligence at the university and information about *who* is doing *what* was important to fulfill this aim.

The cluster map was used to support activities conducted by the project leader, such as interviews and information gathering. It was also used to get overview and display of AI research at KI. Interpretation was found to be challenging in some cases. The interactivity of the map facilitated interpretation. This study was small in scope, but it provides one piece of knowledge about the information need related to algorithmic classifications.

## 1. Introduction

Citation networks have been used since the 1960s to analyze research publications, for example to explore how research fields evolve and how they are delineated into sub-fields (de Solla Price, 1965; Garfield, 1963; Garfield et al., 1964). In the recent decade, large scale classifications of research publications have been created by community detection in citation networks (see e.g. Boyack, 2017; Boyack & Klavans, 2014; Sjögårde & Ahlgren, 2018; Waltman & van Eck, 2012). I herein refer to such classifications as algorithmically constructed publication-level classifications (ACPLC). ACPLCs can be applied globally to all publications in multidisciplinary citation databases or locally to a restricted set of publications.

ACPLCs have predominately been evaluated quantitatively by comparing the ACPLCs to different baselines created from bibliographic data (see e.g. Ahlgren et al., 2020; Boyack & Klavans, 2010; Klavans & Boyack, 2017; Waltman et al., 2020). Such studies are valuable because they enable comparisons between different approaches: notably the use of different publication similarity measures, the use of different clustering algorithms and the use of different parameter values. They also indicate, at least to some extent, whether the delineation of research publications into classes in the ACPLCs are reasonable. However, it is not known if and in which contexts ACPLCs add value for end users.

A few studies have evaluated ACPLCs by expert judgement (external or internal to the author teams) (Ahlgren & Colliander, 2009; Held et al., 2021; Šubelj et al., 2016; Velden et al., 2017). These more qualitative studies have focused on the extent to which ACPLCs correspond to experts' perception of fields and subfields in their domains. No general conclusions can be made from the studies, because they are few, small in scope and the experts have mostly been internal to the group of authors.

Little is known about the information need of potential users of ACPLCs. Many have pointed out that there is no one classification that can be used for all purposes (Glänzel & Schubert, 2003; Gläser et al., 2017; Klavans & Boyack, 2017; Mai, 2011; Smiraglia & van den Heuvel, 2013; Velden et al., 2017; Waltman et al., 2020; Waltman & van Eck, 2012). The choice of methods must therefore be guided by the intended use of an ACPLC. Nevertheless, there has not been much discussion about possible utilization of ACPLCs and how different ACPLCs correspond to different information needs in research organizations. Furthermore, user studies and case studies are missing.

In this case study I have focused on the use of an ACPLC in a university management setting. I studied how an ACPLC can be used to support the promotion of an emerging research field. I focused on a project at Karolinska institutet (KI) called AI@KI. KI is a large medical university with about 3,000 employed researchers and more than 2,000 PhD students. The aim of AI@KI was to strengthen medical research applying artificial intelligence (AI), both in clinical and preclinical settings. To do so, AI@KI aimed to identify AI research at the university and to improve the internal communication between researchers engaged in medical research using AI.

The aim of the current study was to explore and evaluate the usefulness of an ACPLC to managers in a project supporting research in an emerging research area. I focused on management information needs, interpretability of an ACPLC and how the ACPLC corresponded to the information needs. The users being persons in leadership roles at the university and the project leader of AI@KI.

## 2. Background

The role of academic libraries in higher education institutions (HEI) has changed in the last decades. Traditionally the academic libraries have primarily been focusing on providing information to researchers as input to the research process (Åström & Hansson, 2013). However, during the age of digitalization the role has expanded and now includes aspects related to the publication output, such as collecting the publications published by their parent organization, helping to navigate the landscape of scholarly communication as well as evaluating research by the use of bibliometric methods (Åström & Hansson, 2013; Cox et al., 2019; González-Alcaide & Poveda-Pastor, 2018; Vinyard & Colvin, 2018). The use of bibliometrics for research assessment and evaluation has gained much attention in the scientometric research literature (Aksnes & Taxt, 2004, 2004; Bruin et al., 1993; Hallonsten, 2021; Hammarfelt et al., 2016; Hicks et al., 2015; Liu et al., 2015; Moed et al., 1995; *San Francisco Declaration on Research Assessment: Putting Science into the Assessment of Research*, n.d.; Traag & Waltman, 2019). Nevertheless, how bibliometrics is or can be used to provide descriptive (non-evaluative) information in an information management setting is less explored. In particular, the use of ACPLCs as a source of information has not been thoroughly studied in this context. In this paper I focus primarily on the use of ACPLCs in non-evaluative applications.

The computational capacity of computers and theoretical developments within the field of AI has led to an unpreceded ability by machines to learn and perform tasks that hitherto could only be performed by humans. As a result, applications of AI in biomedicine have increased immensely. This development can be seen as an "ecological change" creating an information need for universities to be able to make sense of its surrounding environment (Choo, 1996). ACPLCs have the potential to provide researchers and research management with information of this kind, for example by mapping a research field and its subfields and showing an organization's position in such map. This is an example of bibliometrics extending beyond evaluation. Such practice can be located to the academic library or to other parts of a HEI.

In this paper I focus on the utilization of an ACPLC as a tool used to respond to an information need related to research management. The explication of the information need is a matter of empirical investigation and was not known beforehand. Nonetheless, some information about the project was known at an early stage of this study. The project was initiated by the president of KI and part of the goal was to "to collect and describe all activities at KI related to artificial intelligence (AI)" (Karolinska Institutet, n.d.). In contrast to being evaluative the goal of the project aimed to "exchange knowledge, increase internal and external collaboration, and strengthen KI's research and education in the field." (Ibid).

## 3. Method

I interviewed three of the managers of the project. All three also hold management positions at the university. Thereby, they mediate the information need of the university related to the development of AI activities. The interviews were semi-structured, following a short questionnaire of open questions. The video conference software Zoom was used for the interviews. The focus in these interviews were on the context and aim of AI@KI as well as the information need related to the project. The interviews took around 30 min each.

Two interviews were conducted with the project leader. The first focused on the context, aim, information need and the role of the project leader (similar to the interviews with the management). The second interview was conducted after providing bibliometric data to the project leader. It focused on the use of the ACPLC to respond to the information need in the project. The second interview was supported by the bibliometric data and the respondent used his screen to explain and support his answers.

All questionnaires used for the interviews are presented in the supplementary material. The interviews were analyzed following the thematic structure. I analyzed each theme (for example the information needs related to the project) separately by listening to relevant parts of each recording. I noted key statements related to the theme, for example "there was no picture of how many researchers are working with, or think they are working with, or say they are working with something that can be called AI or machine learning. There was no map […]" [translated from Swedish]. The results presented in this paper is a summary of these statements.

Bibliometric data were provided to the project manager during the project. The bibliometric data were based on AI publications retrieved from a search query. The search query was elaborated after reviewing previous bibliometric analysis of AI. This review resulted in a list of candidate search terms. The project manager was provided this list. He selected search terms and made some amendments. Some Medical Subject Headings (MeSH) were also added to the search query. The search query is provided in the supplementary material as well as a description of the process to obtain it. The query was delimited to the publication period 2016-2021 and was conducted 1 July 2021. It resulted in 78,519 publications, of which 353 had a KI author address.

The bibliometric data included a list of KI publications retrieved from the search query, publication counts per year for KI and for the whole publication set, a co-occurrence network of MeSH, a co-authorship network of KI researchers, a bipartite co-occurrence network of KI researchers and MeSH and map of clusters in an ACPLC. The cluster map contained the levels of disciplines, specialties and topics and was obtained using the Leiden algorithm in a citation network (Traag et al., 2019). The process to determine granularity levels, create labels and visualize the ACPLC has been described in previous papers (Sjögårde, 2021a; Sjögårde et al., 2021; Sjögårde & Ahlgren, 2018, 2020). The 2021 April update of the NIH Open Citation Collection was used for clustering (Hutchins et al., 2019). The classification is available in figshare (Sjögårde, 2021b) and the map of clusters are available in Github.[1]

## 4. Results

### Context and aim of AI@KI

The respondents describe the situation before AI@KI as scattered, with no overall picture of AI activities at KI, lack of communication between research groups, lack of central support in AI related issues and large differences in competence and stage of development. This situation is common in premature areas. It was recognized that KI needed to gather information and connect researchers to be able to develop a more focused and systematic strategy for AI at the university. The managers of AI@KI emphasized that such strategy includes both research and the process to implement AI methods in clinical practice. KI is closely integrated with the health care of Region Stockholm, in particular the Karolinska University Hospital. The respondents saw a high unfulfilled potential of AI methodologies in the clinical setting for diagnoses, treatment, and prevention. Furthermore, they saw several barriers for AI methodologies to reach the clinical practice, such as juridical issues, skepticism, and lack of technical, analytical, and ethical competences. They considered researchers to be in need of more support to be able to take AI from basic research all the way to clinical practice. The connection between AI and precision medicine was also emphasized by the respondents. Precision medicine is a strategic focus area of KI. It was also stressed that medical research projects received AI related grants to a low extent and that a strategic development of AI was essential for the university to maintain a leading position as a medical university.

The aim of AI@KI was to improve competence, enhance communication between researchers, develop support structures and get an overview of the use of AI at KI. The aims related both to AI implementations in basic research and clinical practice.

The respondents expressed that there was a need to enhance the competence of researchers using AI, for the researchers to better formulate research questions, to choose suitable AI methods, and be more successful when applying for grants. They saw a need of more structured approaches when using AI at the university. There was also a need to increase general awareness of AI in all parts of the university, from students to management.

Enhanced communication between researchers was seen as a means to increase the competence of individual researchers using AI or facing research problems that are suitable to be addressed by AI. Furthermore, collaboration was seen as important to develop more systematic approaches to implement AI and to get synergy effects by joining different competences. The respondents considered better communication to be a way to connect competences in order to develop more and better AI related education at the university as well as avoid redundancy. There was also a need to bring researchers together to be able to run large projects and find resources. They emphasized that the aim is to use AI in clinical practice in a way that brings value to patients.

Support structures were needed to help researchers at various stages in the research process, such as formulation of research questions, choice of AI methods, ethical considerations, judicial and regulatory issues, finding collaborations etc. There was no clear picture of how the support structure would be organized at the university. Some express that a center or core facility might be created as a result of the project, but they point out that no such decision has been made. Another possible path forward is a less centralized support structure, based on communication and networks between different units, but without a centralized organization. The horizontal communication between researchers was emphasized in several interviews and that AI must be integrated into the entire university.

An overview of the use of AI at KI was needed to be able to develop activities related to the other aims of the project. Information about *who* is doing *what* within AI was seen as important to connect researchers with shared interests. This information was also seen as important for researchers in early stage of implementing AI. Such researchers need help from others, for example to get training and find collaborators with more experience. Knowing where to find the right competences was seen as important. Another aspect mentioned is the accessibility to structured data. Information about AI activities was seen as crucial for researchers to be able to share data or to develop standards. Furthermore, information about the AI activities was needed by KI for external communication. It was expressed that there was a need for KI to be prepared in order to take part of large investments related to AI and to participate in and contribute to external collaborations.

### Use of the bibliometric data

Activities within AI@KI had been ongoing before the bibliometric data was provided to the project leader. These activities had mainly focused on interviews with researchers using AI and a seminar series. Researchers were reached by an ad hoc social networking strategy. The project leader had contacted researchers known by the project managers. More researchers had been reached through these researchers and other social contacts. The seminar series were also announced, and a growing number of researchers were added to the email list of the seminar series. The project leader expressed that this strategy was not very structured and was risking being biased towards researchers that were known beforehand. However, a more structured strategy to reach researchers through social networking was out of scope for the project.

---

The bibliometric data was used to complement the picture that had arisen from the interviews and the seminars. This was principally done in two ways:

(1) The bibliometric data were used to *support the qualitative* work and to find new areas of AI research to be explored. The project leader used the bibliometric data to find new persons to interview that were not known from previous work. He also found new articles that were relevant to his work in the project, even articles written by interviewed researchers that had not been mentioned. Furthermore, the author networks (including links to each researcher's AI publications) complemented the picture given by the respondents of their own research. The cluster maps showed new, complementing groupings of the AI activities at KI. This use of the bibliometric data was integrated with the qualitative work and affected this work.

(2) In addition, the bibliometric data was used to *get an overview* of AI research at KI. It made it possible to display "the ecosystem of AI research" at KI and give a "taxonomy" of this system. This overview was considered to be broader and more comprehensive. For example, the project leader (and several managers) expressed that there was a common perception that AI in medicine is mostly about using AI for image interpretation, for example in cancer imaging and diagnoses or prediction of cognitive impairment from MRI. However, the cluster map showed that AI at KI is much broader and diversified, for example that AI is extensively used for textual and numerical data besides imaging. Thus, the cluster map showed a new picture of the AI research at KI. Furthermore, it was seen as more systematic and providing evidence of the use of AI at KI.

The interpretation of the cluster map took some effort and was sometimes challenging. The project leader found the labels to be helpful, but not always enough for interpreting the contents of a cluster. The hierarchy and the zooming capability helped the respondent to relate a cluster to its parent, siblings, and children. This information was helpful in interpreting the contents of the cluster. It was rather easy for the respondent to interpret the areas that he was more familiar with than areas which he was less familiar with. He also underlined that he found the relations between areas as interesting and useful.

The grouping of publications into clusters appeared as logical to the respondent. Nevertheless, he stressed that exploration was needed to interpret the clusters. A relatively large cluster of computational biology was mentioned as an example. To understand the contents of this cluster he had to navigate between the underlying specialties, reading labels, the additional provided terms, and the lists of underlying topics. Another example was the division of psychiatry into two different areas (labeled "alzheimer disease; dementia; psychiatry" and "psychiatry; public health; suicide"). Some exploration was needed for the respondent to interpret how the clusters differ. He found this division to be logical, but considered the areas to be overlapping.

## Conclusions

I have studied users' information needs in a project aiming to strengthen collaboration and competence of AI related activities at KI. An important information need in the project was to get knowledge about *who* is doing *what* within AI. This information was essential to build efficient support structures, improve communication and collaboration and increase competence. The overall goal of AI@KI was to develop a more focused, structured, strategic and professional use of AI at the university.

The case study exemplifies how ACPLCs may support managers at HEI by providing information about the topical structure of research output that reflects research activities. The ACPLC based cluster map was useful in the project to completement other activities, including supporting qualitative work, such as interviews and social network building. Furthermore, it provided overview and displayed the research of interest. The ACPLC led to a wider perspective on AI at the university and helped the project to reach researchers in a broader range of research fields.

The study showed that it is not always easy to interpret the contents of clusters. Interpretation of clusters was supported by the possibility to get both overview and detail in the visualization of the ACPLC. The possibility to navigate from broad disciplines down to narrow topics and retrieve individual publications helped the project leader both to understand topics and to find articles of interest. Thereby, this navigation improved the utility of the ACPLC.

This study has provided some insights into user needs and applicability of an ACPLC in practice. To my best knowledge this is the first study of this kind. Knowledge about users' information needs is essential to evaluate and improve the usefulness of ACPLCs in a research management setting. Because of the low number of participants, no general conclusions can be made regarding the usefulness of the ACPLC. The project leader regarded the clusters as making intuitive sense, but sometimes hard to interpret without subject knowledge. Further user studies are needed to evaluate the usefulness of ACPLCs and the interpretability of clusters in relation to information needs in HEI.

It is common knowledge within bibliometrics that coverage of research fields differs in citation databases and that bibliometrics should be used carefully and soundly for this reason. This also applies to descriptive bibliometrics used in this study. However, the study suggests that the bibliometric data could provide a broader and more objective picture of the research within AI than ad hoc social networking.

I suggest future studies to further explore the information needs within HEI that are related to descriptive bibliometrics such as subject mapping. The study suggested that maps incorporating the connection between researchers and research fields would be useful in projects such as AI@KI. Interpretability could also be facilitated by expressing relations between clusters more clearly and if overlap could be visualized. Yet, visualizations must balance between realistic complexity on the one hand and simplification and interpretability on the other.

## ACKNOWLEDGMENTS

I would like to express my gratitude to the respondents.

## FUNDING INFORMATION

Peter Sjögårde was funded by The Foundation for Promotion and Development of Research at Karolinska Institutet.

## Appendix 1: Search strategy

To identify publications in artificial intelligence (AI), I made a review of previous bibliometric studies of AI. I searched Web of Science using the following search query: (TI=bibliometric* OR AB=bibliometric*) AND (TI="artificial intelligence" OR AB="artificial intelligence"). The query was refined to the years 2016 to 2021 and resulted in 68 publications. I read the titles and, if necessary to assess the relevance of the paper, the abstracts of the retrieved publications. I filtered out relevant publications reporting on bibliometric studies of AI, resulting in 20 publications. The full text of 18 of these 20 publications could be accessed. These were downloaded and I identified the methods used to identify AI publications.

Some of the studies have used a set of journals/proceedings to retrieve AI publications (e.g. see Di Vaio et al., 2020; Heradio et al., 2020; Luo et al., 2018; Shukla et al., 2019). Others have used a very simple search strategy, only including the search term "artificial intelligence" (e.g. see Guo et al., 2020; Lei & Liu, 2019; Ruiz-Real et al., 2020; Zhao et al., 2020). "Artificial intelligence" is a broad term and publications using specific AI methods (e.g. "random forest") might not mention "artificial intelligence" and will thereby not be retrieved by such search strategy. Some studies have extended the search to a few more terms (e.g. see Munim et al., 2020; Song & Wang, 2020; Tran et al., 2019). A far more extensive search approach was taken by Chen et al. (2020). They compiled a list of 313 search terms by the following procedure: They first asked domain experts for a list of seed keywords. This list was used to retrieve a set of publications. The most cited publications (10th percentile) were harvested for more keywords. The extended list was manually modified by the authors and then by the domain expert.

I used the list compiled by Chen et al. (2020) as a basis for the search query. The list was reviewed by an AI expert and by the author of this paper (having knowledge of bibliometric methods and search strategies). A large proportion of the search terms were regarded as too unspecific to AI (for example "big data", "information retrieval" and "learning system"). It is possible that these terms were specific enough in the context of the study by Chen et al. However, in this general context such terms would lower the precision of the search query. Terms retrieving less than five hits were also excluded, as well as some redundant terms. The search terms were modified to capture plural by the use of truncation (algorithm*). Finally, four of the Medical Subject Headings under "Artificial Intelligence" were added to the list of search terms ("neural networks, computer", "machine learning", "natural language processing", "computer heuristics"), using expansion to capture any underlying terms. However, four of the terms underlying "Artificial Intelligence" were deemed not to be AI specific ("Robotics", "Expert Systems", "Fuzzy Logic", "Knowledge Bases"), these were excluded to avoid lowering the precision.

**Final list of search terms:**
"artificial bee colony algorithm*"[tiab]

"artificial fish swarm algorithm*"[tiab]
"artificial general intelligence"[tiab]
"artificial intelligence"[tiab]
"artificial neural-network*"[tiab]
"automated inference"[tiab]
"automated reasoning"[tiab]
"big data learning"[tiab]
"brain-machine interface"[tiab]
"conditional random field*"[tiab]
"convolutional neural network"[tiab]
"deep belief network*"[tiab]
"deep learning"[tiab]
"deep network*"[tiab]
"deep neural network"[tiab]
"ensemble learning"[tiab]
"evolutionary computation*"[tiab]
"feature learning"[tiab]
"fuzzy c-means clustering*"[tiab]
"genetic algorithm*"[tiab]
"gradient boosting machine*"[tiab]
"graph embedding*"[tiab]
"kernel method*"[tiab]
"k-means algorithm*"[tiab]
"k-means clustering*"[tiab]
"k-nearest neighbor*"[tiab]
"k-nearest neighbour*"[tiab]
"learning machine*"[tiab]
"machine intelligence"[tiab]
"machine learning"[tiab]
"machine perception"[tiab]
"markov random field*"[tiab]
"multiclass svm"[tiab]
"multiple kernel learning"[tiab]
"natural language processing"[tiab]
"natural language understanding"[tiab]
"nearest neighbor*"[tiab]
"nearest neighbour*"[tiab]
"neural networks (computer)"[tiab]
"perceptron*"[tiab]
"random forest*"[tiab]
"recurrent neural network"[tiab]
"reinforcement learning"[tiab]
"representation learning"[tiab]
"semantic technologies"[tiab]
"semantic technology"[tiab]
"semi-supervised learning"[tiab]
"sequence learning"[tiab]
"supervised classification*"[tiab]
"supervised learning"[tiab]
"supervised machine learning"[tiab]
"support vector classification*"[tiab]
"support vector machine*"[tiab]
"swarm intelligence"[tiab]
"transfer learning"[tiab]

"unsupervised clustering*"[tiab]
"unsupervised learning"[tiab]
"unsupervised machine learning"[tiab]
"neural networks, computer"[mesh]
"machine learning"[mesh]
"natural language processing"[mesh]
"computer heuristics"[mesh]

## Appendix 2: Questionnaires

## Questionnaire for managers

Background, context and purpose of AI@KI
- Can you tell me about the background of the project AI@KI?
- Why was this project started?
- What is the intended project outcome?

Information needs
- According to the web page the aim of AI@KI is to identify and describe activities related to AI at KI. What information is currently missing?
- Why is this information important to collect?
- How is the information going to be used?

Communication
- Will the results of the project be communicated in any way?
- Is there a communication need within the project?

Next step
- How are the results of the project intended to be used?
- How do you imagine the future of AI at KI? Is there a vision?

Other
- Is there anything you want to add?

## Questionnaire for project manager – Interview 1

Role in the project
- Can you describe your view of the background of AI@KI?
- How do you perceive the purpose and aim of the project?
- What is your role in the project?
- What are you expected to deliver?

Information need
- According to the web page the aim of AI@KI is to identify and describe activities related to AI at KI. What information is currently missing?
- What information is needed for you as project leader to fulfill your task?
- How is that information planned to be used?
- What are the challenges to collect the information?
- How have you hitherto acted to collect information?
- Is there any information missing at this point?

Communication
- Will the results of the project be communicated in any way?
- Is there a communication need within the project?

About the bibliometric material
- What information needs do you think can be supported by a bibliometric data?
- In what part of the process is this material needed?
- How can bibliometric data be used?

Other
- Is there anything you want to add?

## Questionnaire for project manager – Interview 2

Questions related to all the delivered bibliometric material
- How has the bibliometric material been used?
  - Has it helped you in the project? How? Which material?

- o Has the bibliometric material helped you to get an overview of AI at KI? Which material?
- o Has the material caused you to act in any way?
- o What has the bibliometric material shown that you did not know from interviews, workshops etc.? Which material?

Specific questions about the subject maps based on algorithmic classification

- Can you describe KI's AI research using the subject maps?
- How has it been to interpret the maps?
- Was it easy or difficult to understand the content of clusters *from the labels*?
- Was it easy or difficult to understand the content of clusters?
- Has it been time consuming to interpret the map?
- If you compare the map with a list of publications, what are the advantages/disadvantages with the map?
- Does the classification show what you have expected?
- How does the functionality work?
- What are the advantages of the map?
- What are the disadvantages?
- Has the map corresponded to your expectations?
- What has been challenging?
- What can be improved?
- Would some other material be more useful?

Other

- Is there anything you want to add?